\documentclass[11pt]{article}
\newcommand{\Conjugate}[1]{ {#1}^{\ast} }
\newcommand{\DiracGamma}[1]{{\gamma}_{#1}}
\newcommand{\Slash}[1]{\rlap/#1}

\newcommand{\VduEb}[2]{V_{d c \tilde{l}^{#1}_{#2} }^{R}}

\newcommand{\VduEd}[2]{V_{d t \tilde{l}^{#1}_{#2} }^{R}}
\newcommand{\Denom}[2]{\frac{1}{#1-#2^2}}
\newcommand{\Denoma}{{\cal P}_1}
\newcommand{\Denomb}{{\cal P}_2}
\newcommand{\Denomc}{{\cal P}_3}
\newcommand{\Denomd}{{\cal P}_4}
\newcommand{\Denome}{{\cal P}_5}

\begin{document}
\begin{large}
\begin{titlepage}

\vspace{0.2cm}

\title{Probing Top-Charm Associated Production at the LHC
              in the $R$-parity violating MSSM
\footnote{The project supported by National Natural Science
          Foundation of China}}
\author{{ Zhou Hong$^{b}$, Ma Wen-Gan$^{a,b}$, Jiang Yi$^{b}$, Zhang Ren-You$^{b}$
          and Wan Lang-Hui$^{b}$}\\
{\small $^{a}$CCAST (World Laboratory), P.O.Box 8730, Beijing
100080, China.}\\
{\small $^{b}$Department of Modern Physics, University of Science
        and Technology}\\
{\small of China (USTC), Hefei, Anhui 230027, China.} }
\date{}
\maketitle

\vskip 12mm

\begin{center}\begin{minipage}{5in}

\vskip 5mm
\begin{center} {\bf Abstract}\end{center}
\baselineskip 0.3in {We present the analytical and numerical
investigations of top-charm associated production at the LHC in
the framework of the $R$-parity violating MSSM. The numerical
analysis of their production rates is carried out in the mSUGRA
scenario with some typical parameter sets. The results show that
the cross sections of associated $t\bar{c}(\bar{t}c)$ production
via gluon-gluon fusion can reach $5\%$ of that via $d\bar{d}$
annihilation. The total cross section will reach the order of $10
\sim 10^2 ~fb$ and the cross sections are strongly related to the
$R$-parity violating parameters. }\\
\vskip 5mm {PACS number(s): 12.60.Jv, 14.65.-q, 14.65.Ha,
14.65.Dw}
\end{minipage}
\end{center}
\end{titlepage}

\baselineskip=0.36in

\eject
\rm
\baselineskip=0.36in

\begin{flushleft} {\bf 1. INTRODUCTION} \end{flushleft}
\par
There are stringent experimental constraints against the existence
of tree-level flavor changing scalar interactions(FCSI's)
involving the light quarks. This leads to the suppression of the
flavor changing neutral current (FCNC) couplings, an important
feature of the standard model (SM), which is explained in terms of
the Glashow-Iliopoulos-Maiani(GIM) mechanism \cite{Glashow}. At
present, the minimal supersymmetric extension (MSSM)\cite{haber}
\cite{Gunion} of the standard model (SM) \cite{glash}\cite{higgs}
is widely considered as the most appealing model. Apart from
describing the experimental data as well as the SM does, the
supersymmetric (SUSY) theory is able to solve various theoretical
problems, such as the fact that the SUSY may provide an elegant
way to construct the huge hierarchy between the electroweak
symmetry-breaking and the grand unification scales.
\par
FCNC coupling is widely studied for its importance to verify new
physics. Searching for FCNC at high energy colliders, particularly
$e^+e^-$ colliders was investigated in Ref.\cite{FCNC1}. Probing
the FCNC vertices $\bar{t}$-c-V(V=$\gamma$, Z) in rare decays of
top quark and via top-charm associated production were examined in
Refs.\cite{FCNC2} and \cite{FCNC3}-\cite{Yu1}, respectively. The
effect of the anomalous $\bar{t}-c-g$ coupling on single top quark
production via the $q \bar{q}$ process at the Tevatron has been
studied in Ref.\cite{Malkawix}, Here we mention some possible
mechanisms which can induce the FCNC couplings:\\
1. In the Standard Model (SM), the FCNC couplings are strongly
suppressed by GIM mechanism. Such interactions can be produced by
higher order radiative corrections in the SM, the effect is too
small to be observable \cite{FCNC3} \cite{Eilamx}.\\
2. In models with multiple Higgs doublets such as supersymmetric
models and the Two-Higgs-Doublet-Model(THDM) (model III), there
would exist possible strong effects of the FCNC \cite{Eilamx}
\cite{Grzadx}. Atwood et al. \cite{atwood}\cite{lin} presented the
results of a calculation for the process $e^{+}e^{-}\rightarrow t
\bar{c}$(or $\bar{t}c$) in the THDM III. In Ref.\cite{lin}
\cite{Jiang} \cite{li}, the process $\gamma \gamma \rightarrow
t\bar{c}$(or $\bar{t}c$) in the THDM III and SUSY-QCD, is studied
at the Next Linear Collider. The associated product of
$t\bar{c}(\bar{t}c)$ via gluon-gluon at hadron colliders was
consider by \cite{Chang}. They all concluded that it would be
possible to find associated $t \bar{c}$(or $\bar{t}c$) production
events at the NLC, Tevatron and LHC in the THDM (III) and the
MSSM. They also showed that the FCNC effects depended on the
resonance of Higgs boson. In the MSSM with $R$-parity
conservation, squark mixing can give FCNC couplings. But if we
take alignment assumption of S. Dimopoulos \cite{Dimopoulos},it
should be very small: mixing between up-type squarks can be even
as small as $10^{-3}$ to $10^{-5}$ times KM matrix elements.
\par
In the MSSM, if lepton and baryon numbers are conserved, there
must be a conservation of a discrete symmetry called
R-parity($R_{p}$) conservation\cite{Farrar}, which is defined as
$$
R_{p}=(-1)^{3B+L+2S},
$$
where B, L and S are the baryon, lepton number and spin of a
particle, respectively. In this case, all supersymmetric particles
must produced in pair, and the lightest supersymmetric particle
must be stable.
\par
However, $R_{p}$ conservation with both B- and L-number conserved
is not necessary to avoid rapid proton decays, instead we just
need either B-conservation or L-conservation\cite{Hall}. In this
case the R-parity is not conserved any more and the feature of
supersymmetric models are changed a lot. Due to the lack of
experimental tests for $R_{p}$ conservation, the $R_{p}$ violation
is also equally well motivated in the MSSM. And the models with
$R_{p}$ violation $(\rlap/R_{p})$ are hopeful for us to solve the
long standing problems in the particle physics, such as neutrino
masses and mixing.
\par
Theoretically $R_{p}$-violation models will open some new
processes forbidden or highly suppressed in $R_{p}$ conservation
case, but the present low-energy experimental data have put
constraints on $R_p$-violation parameters. Unfortunately, they
give only some upper limits on the $\rlap/R_{p}$ parameters, such
as B-violating parameters( $\lambda^{"}$) and L-violating
parameters($\lambda$ and $\lambda^{'}$) (The definitions of these
$\rlap/R_{p}$ parameters will be presented clearly in sector 2,
and their constraints are collected in Ref.\cite{Allanach}.).
Therefore, trying to find the signal of $R_{p}$ violation or
getting more stringent constraints on the parameters in future
experiments is one of the promising tasks.
\par
In the last few years, many efforts were made to find
$\rlap/R_{p}$ interactions in experiments. The possible signal of
$R_{p}$-violation could be the single SUSY particle production or
LSP decay, the existence of the difference between the fermion
pair production rates in the $\rlap/R_{p}$ MSSM and $R_{p}$
conservation MSSM, and probing couplings of the flavor changing
neutral current(FCNC) et cetera.
\par
In the following years, the hadron colliders, such as Tevatron Run
II and the LHC, are the effective machines in searching for new
physics. People believe that there will be more experimental
events involving top quark collected in the future experiments. It
provides an opportunity to study the physics beyond the SM with
more precise experimental results.
\par
In this work we will concentrate on the FCNC coupling test and use
associated $\bar{t}$c(or t$\bar{c}$) production at LHC to probe
$R_{p}$ violation. Although up to now many constraints from
low-energy phenomenology have been given, B-violation parameters
involving heavy flavors are still constrained weakly. Such as
$\lambda^{"}_{2ij}$ and $\lambda^{"}_{3ij}$, which got strongest
constraints from width ratio between $Z^0$ decaying to leptons and
hadrons, can still be order of 1(O(1)). So if these parameters are
standing close to present upper limits, $R_p$-violating effects
could be detected on future colliders.
\par
In this paper we present the complete parent process
$pp\rightarrow t\bar{c}(\bar{t}c)$ including one-loop induced
subprocess $gg \rightarrow t\bar{c}(\bar{t}c)$ and tree-level
subprocess $dd \rightarrow t\bar{c}(\bar{t}c)$ in the R-parity
violating MSSM theory. The paper is arranged as follows: In Sec.2
we give the analytical calculations of both subprocess and parent
process. In Sec.3, the numerical results for subprocess and parent
process are illustrated along with discussions. A short summary is
presented in Sec.4 . Finally some notations used in this paper,
the explicit expressions of the form factors induced by the loop
diagrams are collected in Appendix.

\begin{flushleft} {\bf 2. CALCULATION} \end{flushleft}
\par
The $R_p$ violating MSSM should contain the most general
superpotential respecting to the gauge symmetries of the SM, which
includes bilinear and trilinear terms and can be expressed as
\begin{equation}
\label{sup}
    {\cal W}_{\rlap/R_{p}} =\frac{1}{2}
\lambda_{[ij]k} L_{i}.L_{j}\bar{E}_{k}+\lambda^{'}_{ijk}
L_{i}.Q_{j}\bar{D_{k}}+\frac{1}{2}\lambda^{''}_{i[jk]}
\bar{U}_{i}\bar{D}_{j}\bar{D}_{k}+\epsilon _{i} L_{i} H_{u}.
\end{equation}
where $L_i$, $Q_i$ are the SU(2) doublet lepton and quark fields,
$E_i$, $U_i$, $D_i$ are the singlet superfields. The $UDD$
couplings violate baryon number and the other three sets violate
lepton number. In this work we ignored the bilinear term that
includes lepton and Higgs superfields  for simplicity, because its
effects are assumed small in our process\cite{Hall}. We also forbid
explicitly the $UDD$-type interactions (B-number violation) as a
simple way to avoid unacceptable rapid proton decay\cite{Ross}.
Since the couplings in the term of $LLE$ have no contribution to
the process $pp \rightarrow t\bar{c}(\bar{t}c)+X$ concerned in
this paper, we shall not discuss them either.
\par
Expanding the second term of superfield components in
Eq.(\ref{sup}) we obtain the interaction Lagrangian that involves
quarks and leptons:
\begin{equation}
\label{lag}
{\cal L}_{LQD}=\lambda_{ijk}^{'}\{\tilde{\nu}_{iL}\bar{d}_{kR}d_{jL}-
      \tilde{e}_{iL}\bar{d}_{kR}u_{jL}+\tilde{d}_{jL}
      \bar{d}_{kR}\nu_{iL}- \tilde{u}_{jL}\bar{d}_{kR}e_{iL}+
      \tilde{d}_{kR}^c\nu_{iL}d_{jL}- \tilde{d}_{kR}^ce_{iL}u_{jL}\} + h.c.
\end{equation}
The Feynman diagrams contributing to the tree-level subprocess
$d\bar{d} \rightarrow t\bar{c}(\bar{t}c) $ in the framework of the
$\rlap/{R}_{p}$-MSSM is depicted in Fig.1(tree-level). In our
calculation, we take the 't Hooft-Feynman gauge. The related
Feynman rules with $\rlap/{R}_{p}$ interactions can be read out
from Eq.(\ref{lag}). In the following we adopt the notations in
Ref.\cite{wanlh} that $p_1$ and $p_2$ represent the four-momenta
of the incoming particles and $k_1$ and $k_2$ represent the
four-momenta of the outgoing quarks t and $\bar{c}$ respectively.
If we ignore the CP violation, the cross section of $pp
\rightarrow d\bar{d} \rightarrow t\bar{c} +X$ coincides with the
process $pp \rightarrow d\bar{d} \rightarrow \bar{t}c +X$ because
of charge conjugation invariance, and the same is also for the
loop process $pp \rightarrow gg \rightarrow t\bar{c} +X$.
Therefore, we shall consider only the calculation of the
$t\bar{c}$ production in this paper. The corresponding
Lorentz-invariant matrix element at the lowest order for the
subprocess $d\bar{d} \rightarrow t\bar{c}$ is written as
$$
{\cal M}(d\bar{d} \rightarrow
t\bar{c})=\sum\limits_{\tilde{l}_{i}^{I}} {\cal
M}_{\tilde{l}_{i}^{I}}
$$
where $\tilde{l}_{i}^{I}$ is the partner of lepton $l^{I}$, $i$ and $I$ are
the mass eigenstate and the generation indeces, respectively.
The corresponding differential cross section is obtained by
$$
\frac{d \hat{\sigma}} {d \Omega}=
   {\frac{\lambda}{64\pi^2\hat{s}^{2}}}\bar{|{\cal M}|}^2
$$
where $\lambda=\sqrt{
\left[\hat{s}-(m_{t}+ m_{c})^2\right]
\left[\hat{s}-(m_{t}- m_{c})^2\right]}$.
\par
For the subprocess of $d\bar{d} \rightarrow t\bar{c}$,
 \begin{eqnarray*}
 && \bar{|{\cal M}|}^2 =\sum\limits_{\tilde{l}_{i}^{I}, ~\tilde{l}_{j}^{J}}
 \Denom{\hat{t}}{m_{\tilde{l}_{i}^{I}}}
\Denom{\hat{t}}{m_{\tilde{l}_{j}^{J}}}
 (k_{1} \cdot p_{1}) (k_{2} \cdot p_{2})(
 \Conjugate{\VduEb{J}{j}}
  \Conjugate{\VduEd{I}{i}} \VduEb{I}{i} \VduEd{J}{j})
\end{eqnarray*}
After integrating over phase space $\Omega$ we can get the total
section of $d\bar{d} \rightarrow t\bar{c}$
\begin{eqnarray*}
\hat{\sigma}(d\bar{d} \rightarrow t\bar{c})&=&  \frac{1}{64\pi
\hat{s}^{2}} \sum\limits_{\tilde{l}_{i}^{I}, ~\tilde{l}_{j}^{J}}
\Conjugate{\VduEb{J}{j}}
  \Conjugate{\VduEd{I}{i}} \VduEb{I}{i} \VduEd{J}{j}  \times \\
&& \left\{\delta_{\tilde{l}_{i}^{I},~\tilde{l}_{j}^{J}}
\left[\lambda \left( 1+ \frac{4 \beta_{\tilde{l}_{i}^{I}}}
{\alpha_{+}\alpha_{-}} \right)
+ (2 m_{\tilde{l}_{i}^{I}}^2 - m_c^2 - m_t^2 ) \gamma_{\tilde{l}_{i}^{I}} \right] \right. \\
&+& \left. (1-\delta_{\tilde{l}_{i}^{I},~\tilde{l}_{j}^{J}})\left( \lambda
 +\frac{\beta_{\tilde{l}_{i}^{I}}\gamma_{\tilde{l}_{i}^{I}}}
 {m_{\tilde{l}_{i}^{I}}^2 - m_{\tilde{l}_{j}^{J}}^2}
 - \frac{\beta_{\tilde{l}_{j}^{J}}\gamma_{\tilde{l}_{j}^{J}}}
   {m_{\tilde{l}_{i}^{I}}^2 - m_{\tilde{l}_{j}^{J}}^2}
   \right) \right \}
\end{eqnarray*}
where we define the notations as
$$
\alpha_{\pm}=m_c^2 + m_t^2- 2 m_{\tilde{l}_{i}^{I}}^2  - \hat{s} \pm \lambda,
$$
$$
\beta_{k}=(m_c^2 - m_k^2)(m_t^2 - m_k^2),
$$
$$
\gamma_{k}=\log \left(\frac{ m_c^2 + m_t^2
      - 2 m_k^2  - \hat{s} + \lambda }{ m_c^2 + m_t^2
  - 2 m_k^2  - \hat{s}-\lambda
  }\right),~~~(k=\tilde{l}_{i}^{I},\tilde{l}_{j}^{J}).
$$
In the above equation, the bars over  $\cal M$ mean
average over initial spin and color.The $\delta$ is the Kronecker
delta. The notations for vertices are adopted which are shown in
Appendix and $\hat{t}=(p_1-k_1)^2$.
\par
 The subprocess $gg \rightarrow
t\bar{c}(\bar{t}c)$ can only be produced through one-loop diagrams
at the lowest order. Due to the large gluon luminosity in protons,
the contribution of one-loop subprocess $gg \rightarrow
t\bar{c}(\bar{t}c)$ to the parent process $pp \rightarrow
t\bar{c}(\bar{t}c)$ can be significant. In the calculation of
subprocess $gg \rightarrow t\bar{c} (\bar{t}c)$, it is not
necessary to consider the renormalization, since the ultraviolet
divergence will be cancelled automatically when all the one-loop
diagrams in framework of the $R_p$-violating MSSM are involved.
The generic Feynman diagrams of the subprocess are depicted in
Fig.1(1-31), where the possible exchange of incoming gluons in
Fig.1b are not shown. We denote the reaction of $t\bar{c}$
production via gluon-gluon fusion as:
\begin{equation}
g(p_1, \alpha, \mu) g(p_2,{\alpha}^{'} , \nu) \longrightarrow
t(k_1,\beta ) \bar{c}(k_2,{\beta}^{'}).
\end{equation}
where $p_1$ and $p_2$ denote the four momenta of the incoming gluons,
$k_1$, $k_2$ denote the four momenta of the outgoing t and $\bar{c}$
respectively, and $\alpha$,${\alpha}^{'}$  are the color indices of the colliding
gluons; $\beta$, ${\beta}^{'}$ are the color indices of the produced particles.
\par
The corresponding matrix element of the subprocess $gg \rightarrow
t\bar{c}(\bar{t}c)$ can be divided into four parts:
\begin{equation}
{\cal M}={\cal M}^{\hat{t}}+{\cal M}^{\hat{u}}+{\cal M}^{\hat{s}}+{\cal M}^{q}
\end{equation}
${\cal M}^{q}$ is the amplitude of quartic diagram. The $u$-channel part can be
obtained from the $t$-channel part by doing exchanges as shown below:
\begin{equation}
{\cal M}^{\hat{u}} = {\cal\ M}^{\hat t}(\hat{t}\rightarrow \hat{u}, k_1 \leftrightarrow k_2,
              \mu \leftrightarrow \nu,\alpha \leftrightarrow {\alpha}^{'})
\end{equation}
\par
The corresponding matrix element of the subprocess $gg \rightarrow
t\bar{c}$ for $\hat{t}$-channel, s-channel and quartic interaction
diagrams shown in Fig.1b can be written as:
\begin{eqnarray*}
{\cal M}^{\hat{t}}&=&\epsilon^{\mu}(p_{1})\epsilon^{\nu}(p_{2})
\bar{u}(k_1) \{ f_{1}^{\hat{t}} g_{\mu \nu} +
 f_{2}^{\hat{t}} \DiracGamma{\mu}  \DiracGamma{\nu} +
 f_{3}^{\hat{t}} k_{1\mu} k_{1\nu} +
 f_{4}^{\hat{t}} \DiracGamma{\nu} k_{1\mu} +
 f_{5}^{\hat{t}} \DiracGamma{\mu} k_{1\nu}\\ &+&
 f_{6}^{\hat{t}} g_{\mu \nu} \Slash{p_{1}} +
 f_{7}^{\hat{t}} \DiracGamma{\mu}  \DiracGamma{\nu}  \Slash{p_{1}} +
 f_{8}^{\hat{t}} k_{1\mu} k_{1\nu} \Slash{p_{1}}+
 f_{9}^{\hat{t}} k_{1\mu} \DiracGamma{\nu}  \Slash{p_{1}} +
 f_{10}^{\hat{t}} k_{1\nu} \DiracGamma{\mu}  \Slash{p_{1}} \\ &+&
 f_{11}^{\hat{t}} \DiracGamma{5} g_{\mu \nu} +
 f_{12}^{\hat{t}} \DiracGamma{5}  \DiracGamma{\mu}  \DiracGamma{\nu} +
 f_{13}^{\hat{t}} \DiracGamma{5}  k_{1\mu} k_{1\nu}  +
 f_{14}^{\hat{t}} k_{1\mu} \DiracGamma{5}  \DiracGamma{\nu} +
 f_{15}^{\hat{t}} k_{1\nu} \DiracGamma{5}  \DiracGamma{\mu} \\ &+&
 f_{16}^{\hat{t}} g_{\mu \nu} \DiracGamma{5}  \Slash{p_{1}} +
 f_{17}^{\hat{t}} \DiracGamma{5}  \DiracGamma{\mu}  \DiracGamma{\nu}  \Slash{p_{1}} +
 f_{18}^{\hat{t}} k_{1\mu} k_{1\nu} \DiracGamma{5}  \Slash{p_{1}} +
 f_{19}^{\hat{t}} k_{1\mu} \DiracGamma{5}  \DiracGamma{\nu}  \Slash{p_{1}} +
 f_{20}^{\hat{t}} k_{1\nu} \DiracGamma{5}  \DiracGamma{\mu}  \Slash{p_{1}}
\}\\
&&v(k_2) T^{\alpha}_{\beta c} T^{{\alpha}^{'}}_{c{\beta}^{'}} \\
{\cal M}^{\hat{s}}&=&\epsilon^{\mu}(p_{1})\epsilon^{\nu}(p_{2})
\bar{u}(k_1) \{ f_{1}^{\hat{s}} g_{\mu \nu} +
f_{6}^{\hat{s}} g_{\mu \nu} \Slash{p_{1}} +
f_{11}^{\hat{s}} \DiracGamma{5} g_{\mu \nu} +
f_{16}^{\hat{s}} g_{\mu \nu} \DiracGamma{5}  \Slash{p_{1}}
\}
v(k_2)( T^{\alpha}_{\beta c} T^{{\alpha}^{'}}_{c{\beta}^{'}}
       -T^{{\alpha}^{'}}_{\beta c} T^{\alpha}_{c{\beta}^{'}})\\
{\cal M}^{q}&=&\epsilon^{\mu}(p_{1})\epsilon^{\nu}(p_{2})
\bar{u}(k_1) \{ f_{1}^{q} g_{\mu \nu} + f_{11}^{q} \DiracGamma{5} g_{\mu \nu} \}
v(k_2)( T^{\alpha}_{\beta c} T^{{\alpha}^{'}}_{c{\beta}^{'}}
       +T^{{\alpha}^{'}}_{\beta c} T^{\alpha}_{c{\beta}^{'}})\\
\end{eqnarray*}
where $T^{a}_{i j}$ are the $3 \times 3$ SU(3) color matrices
introduce by Gell-Mann \cite{Gell}. We divide each form factor $f_{i}^{\hat{t}}$
into follows
\begin{eqnarray*}
f_{i}^{\hat{t}}=f_{i}^{b,\hat{t}}+f_{i}^{v,\hat{t}}+f_{i}^{s,\hat{t}}~~(i=1-20)
\end{eqnarray*}
\par
The explicit expressions of form factors are collected in Appendix.
The cross section for this subprocess at one loop order via unpolarized
gluon collisions can be got by using the following equation,
\begin{equation}
\label{sss}
\hat{\sigma}(\hat{s},gg \rightarrow t\bar{c}) =
       \frac{1}{16 \pi \hat{s}^2} \int_{\hat{t}^{-}}^{\hat{t}^{+}}
       d\hat{t}~ \bar{\sum\limits_{}^{}} |{\cal M}|^2.
\end{equation}
In above equation, $\hat{t}$ is the momentum transfer squared from one
of the incoming gluons to the quark in the final state, and
$$
\hat{t}^\pm=\frac{1}{2}\left[ (m^{2}_{t}+m^{2}_{c}
-\hat{s})\pm \sqrt{(m^{2}_{t}+m^{2}_{c}-\hat{s})^2
-4m^{2}_{t}m^{2}_{c}} \right].
$$
The bar over the sum means average over initial spin and color.
With the results from Eq.(\ref{sss}), we can easily obtain the
total cross section at $pp$ collider by folding the cross section
of subprocess $\hat{\sigma} (gg \rightarrow t\bar{c})$ with the
gluon luminosity.

\begin{equation}
\sigma(s,pp \rightarrow gg \rightarrow  t\bar{c}+X)=
   \int_{(m_{t}+m_{c})^2/
s} ^{1} d \tau \frac{d{\cal L}_{gg}}{d \tau} \hat{\sigma}
(gg\rightarrow t\bar{c}
\hskip 3mm at \hskip 3mm \hat{s}=\tau s),
\end{equation}

where $\sqrt{s}$ and $\sqrt{\hat{s}}$ are the $pp$ and $gg$
c.m.s. energies respectively and $d{\cal L}_{gg}/d \tau$
is the distribution function of gluon luminosity, which is defined as
\begin{equation}
\frac{d{\cal L}_{gg}}{d\tau}=\int_{\tau}^{1}
\frac{dx_1}{x_1} \left[ f_g(x_1,Q^2)f_g(\frac{\tau}{x_1},Q^2) \right].
\end{equation}
here $\tau = x_1~x_2$, the definition of $x_1$ and $x_2$ are from
\cite{jiang}, and in our calculation we adopt the MRS set G parton
distribution function \cite{Martin}. The factorization scale Q was
chosen as the average of the final particles masses
$\frac{1}{2}(m_{t} +m_{c})$. The total cross section contributed
by the subprocess $d \bar{d} \rightarrow t\bar{c}(\bar{t}c)$ can
be obtained by the same way claimed above. The total cross section
of $pp \rightarrow t\bar{c}+\bar{t}c+X$ is obtained by the cross
section of $pp \rightarrow t\bar{c}+X$ multiplied by factor 2.

\par
\begin{flushleft} {\bf 3. Numerical results and discussions} \end{flushleft}
\par
In the following numerical evaluation, we present the numerical
results of the cross sections for the $t\bar{c}(\bar{t}c)$
production in the subprocesses and parent process. The parameters
originating from the SM are chosen as: quark and lepton mass
parameters are obtained from Ref.\cite{booklet}. We take a simple
one-loop formula for the running strong coupling constant
$\alpha_s$. We set $\alpha_s(m_Z)=0.117$ and $n_f=5$.
\par
The R-parity violating parameters involvd in the evaluation are
set to be
${\lambda}^{'}_{1ij}={\lambda}^{'}_{2ij}={\lambda}^{'}_{3ij}=0.15
$ unless otherwise stated explicitly. As we know that the effects of the R-parity violating couplings
on the renormalization group equations(RGE's) are the crucial
ingredient of mSUGRA-type models, and the complete 2-loop RGE's of
the superpotential parameters for the supersymmetric standard
model including the full set of R-parity violating couplings are
given in Ref.\cite{Allanach}. But in our numerical presentation to
get the low energy scenario from the mSUGRA \cite{Drees}, we
ignored those effects in the RGE's for simplicity and use the
program ISAJET 7.44. In this program the RGE's \cite{RGE} are run
from the weak scale $m_{Z}$ up to the GUT scale, taking all
thresholds into account and using two loop RGE's only for the
gauge couplings and the one-loop RGE's for the other
supersymmetric parameters. The GUT scale boundary conditions are
imposed and the RGE's are run back to $m_Z$, again taking
threshold into account.The R-parity violating parameters chosen
above satisfy the constraints given by \cite{Hall}.
\par
Figure 2 shows the cross sections as a function of
$\sqrt{\hat{s}}$, and the upper curve corresponds to the
subprocess $d\bar{d} \rightarrow t\bar{c}$ and the lower curve
corresponds to the subprocess $gg \rightarrow t\bar{c}$. The input
parameters are chosen as $m_{0}=180~GeV,
m_{\frac{1}{2}}=150~GeV,A_{0}=200~GeV, \tan\beta=4, sign(\mu)=+$ .
With above parameters, we get $m_{{\tilde{b}}_{1}}=353~GeV,
m_{{\tilde{b}}_{2}}=375~GeV,m_{{\tilde{d}}_{1}}=m_{{\tilde{s}}_{1}}=375~GeV,
m_{{\tilde{d}}_{2}}=m_{{\tilde{s}}_{2}}=390~GeV$ in the framework
of the mSUGRA. Due to the threshold effects, we can see sharp
rising peaks around $\sqrt{\hat{s}} \sim 180~GeV$ on the two
curves in Figure 2, where the threshold condition $\sqrt{\hat{s}}
\sim m_t+m_c$ is satisfied. For the subprocess $gg \rightarrow
t\bar{c}$, when $\sqrt{\hat{s}}$ approaches the value of
$2m_{\tilde{d}}$, the cross section will be enhanced  by the
resonance effects. The small peak on the curve of subprocess $gg
\rightarrow t\bar{c}$, where $\sqrt{\hat{s}} \sim 2m_{\tilde{d}}
\simeq 780~GeV$, comes from the resonant effect of the quartic
diagrams.
\par
The integrated cross sections versus $\tan\beta$ are depicted in
Figure 3 and versus $m_{0}$ in Figure 4, respectively. We
calculate the $t\bar{c}+\bar{t}c$ production cross sections at the
LHC with the energies of $\sqrt{s}$ being $14~TeV$. In Figure 3
the input parameters are chosen as $m_{0}=150~GeV,
m_{\frac{1}{2}}=150~GeV, A_{0}=200~GeV, sign(\mu)=+$, and in
Figure 4 as $m_{\frac{1}{2}}=150~GeV, A_{0}=200~GeV, \tan\beta=4,
sign(\mu)=+$. In both figures, the dotted lines are the curves
contributed by $d\bar{d} \rightarrow t\bar{c}+\bar{t}c$, the
dashed lines are the curves contributed by $gg \rightarrow
t\bar{c}+\bar{t}c$ and the solid lines are the curves of total
cross sections which are the sum of the above two subprocesses.
Usually it is shown that the cross section contribution to parent
process at hadron collider from subprocess $gg \rightarrow
t\bar{c}+\bar{t}c$ can be about $5\%$ of that from subprocess
$d\bar{d} \rightarrow t\bar{c}+\bar{t}c$. So the production
mechanism of subprocess $gg \rightarrow t\bar{c}+\bar{t}c$ should
be considered in detecting the $\rlap/{R}_{p}$ signals in this
parameter space.
\par
In Figure 3 $\tan\beta$ varies from 2 to 30. The total cross
section decreases first and at the position of $\tan\beta \simeq
5$ it arrives the nadir, then it increase slightly. The cross
section via $pp \rightarrow d\bar{d} \rightarrow t\bar{c}$ has the
same feature, but the curve for the cross section via $pp
\rightarrow gg \rightarrow t\bar{c}$ has little different. In the
framework of the mSUGRA, when $m_0$ varies from $180~GeV$ to
$300~GeV$, $m_{\tilde{d}}$ ranges from $370~GeV$ to $440~GeV$. So
we can see in Figure 4 that the cross section decreases rapidly
with the increment of $m_0$.
\par
Finally, we will focus on the relationship between the $\bar{t}c +
t \bar{c}$ production cross section at the LHC and the
$R_p$-violation parameters $\lambda^{'}_{ijk}$. The sensitivity of
the cross section of parents process $ pp \rightarrow d
\bar{d}(gg) \rightarrow \bar{t} c+t \bar{c}$ to
$\lambda^{'}_{331}* \lambda^{'}_{321}$ with other
$\lambda^{'}_{ijk}$'s being taken as 0.15, are shown in Figure 5
in the mSUGRA scenario, where the input parameters $m_{0}$,
$m_{\frac{1}{2}}$, $A_{0}$, $\tan\beta$, $sign(\mu)$ are taken as
the same as the corresponding ones in Figure 2. The dotted line is
the curve contributed by subprocess $d\bar{d} \rightarrow
t\bar{c}+\bar{t}c$, the dashed line is the curve contributed by
$gg \rightarrow t\bar{c}+\bar{t}c$. The cross sections of the both
subprocesses are all the functions of $((\lambda^{'}_{331}*
\lambda^{'}_{321}))^2$. Therefore, the dependence of the
production cross section of $t\bar{c}+\bar{t}c$ on the values of
$\lambda^{'}_{ijk}$ is very strong. In the allowable parameter
space of $\lambda^{'}_{ijk}$ \cite{Allanach}, the cross sections
will cover a great range. Similar with the case of the L-number
violating case, in the B-number violating case, the
$R_p$-violation parameters $\lambda^{"}_{ijk}$ could play
significant role also in the top-charm associated production at
the LHC, but we will not discuss it in details in this paper.

\par
\begin{flushleft} {\bf 4. Summary} \end{flushleft}
\par
In this paper, we have studied the production of top-charm
associated production with explicit $R_{p}$-violation at the LHC.
The production rates via $d-\bar{d}$ annihilation and gluon-gluon
fusion at the LHC are presented analytically and numerically in
the mSUGRA scenario with some typical parameter sets. The results
show that the cross section of the top-charm associated production
at the LHC via gluon-gluon collisions can reach about several
femto barn with our chosen parameters, and is usually about $5\%$
of that via quark-antiquark annihilation subprocess. It means that
the contribution from $gg \rightarrow t\bar{c}(\bar{t}c)$
subprocess can be competitive with that via $d\bar{d} \rightarrow
t\bar{c}(\bar{t}c)$ subprocess at the LHC and can be considered as
an important part of the NLO QCD correction to the $pp \rightarrow
t\bar{c}(\bar{t}c)+X$ subprocess. Therefore, in detecting the
top-charm associated production at the LHC in searching for the
signals of SUSY and $R_{p}$ violation, we should consider not only
the associated $t\bar{c}(\bar{t}c)$ production via quark-antiquark
annihilation, but also that via the gluon-gluon fusion. By taking
an annual luminosity at the LHC being $100~fb^{-1}$, one may
accumulate $10^3$ $t\bar{c}(\bar{t}c)$ production events per year.

{\Large{\bf Appendix}}
\vskip 5mm
\par
The relevant Feynman rules concerned in this work are list below:
\begin{eqnarray*}
\bar{D}-U-\tilde{L_{i}}&&:~~~~
V_{d^{K}u^{J}\tilde{l_{i}}^{I}}^R~ P_R \\
\bar{U}-\bar{L}-\tilde{D_{i}}&&:~~~~
V_{\tilde{d_{i}}^{K}l^{I}u^{J}}^L ~P_L~ C
\end{eqnarray*}
where $C$ is the charge conjugation operator, $P_{L,R} =
\frac{1}{2}(1 \mp \gamma_5)$. The vertices can be read out from
Eq.(\ref{lag}):
\begin{eqnarray*}
V_{d^{K}u^{J}\tilde{l_{1}}^{I}}^R =i \lambda_{IJK}^{'} \cos \theta_{\tilde{L}}&~~&
V_{d^{K}u^{J}\tilde{l_{2}}^{I}}^R =i \lambda_{IJK}^{'} \sin \theta_{\tilde{L}}
\\
V_{\tilde{d_{1}}^{K}l^{I}u^{J}}^L =-i \lambda_{IJK}^{'} \sin \theta_{\tilde{D}}&~~&
V_{\tilde{d_{2}}^{K}l^{I}u^{J}}^L =i \lambda_{IJK}^{'} \cos \theta_{\tilde{D}}
\end{eqnarray*}
\par
We adopt the same definitions of one-loop A, B, C and D integral
functions as in Ref.\cite{abcd} and the references therein. All
the vector and tensor integrals can be deduced in the forms of
scalar integrals \cite{veltman}. The dimension $D=4- \epsilon$.
The integral functions are defined as
$$
A_{0}(m)=-\frac{(2\pi\mu)^{4-D}}{i\pi^{2}} \int d^{D}q \frac{1}{[q^2-m^2]} ,
$$
$$
\{B_{1};B_{\mu};B_{\mu\nu}\}(p,m_1,m_2) =
\frac{(2\pi\mu)^{4-D}}{i\pi^{2}} \int d^{D}q
\frac{\{1;q_{\mu};q_{\mu\nu}\}}{[q^2-m_{1}^{2}][(q+p)^2-m_{2}^{2}]} ,
$$
$$
\{C_{0};C_{\mu};C_{\mu\nu};C_{\mu\nu\rho}\}(p_1,p_2,m_1,m_2,m_3) =
-\frac{(2\pi\mu)^{4-D}}{i\pi^{2}}
$$
$$
\times \int d^{D}q
\frac{\{1;q_{\mu};q_{\mu\nu};q_{\mu\nu\rho}\}}
{[q^2-m_{1}^{2}][(q+p_1)^2-m_{2}^{2}][(q+p_1+p_2)^2-m_{3}^{2}]} ,
$$
$$
\{D_{0};D_{\mu};D_{\mu\nu};D_{\mu\nu\rho};D_{\mu\nu\rho\alpha}\}
(p_1,p_2,p_3,m_1,m_2,m_3,m_4) =
\frac{(2\pi\mu)^{4-D}}{i\pi^{2}}
$$
$$
\times \int d^{D}q \{1;q_{\mu};q_{\mu\nu};q_{\mu\nu\rho};
q_{\mu\nu\rho\alpha}\}
$$
$$
\times \{[q^2-m_{1}^{2}][(q+p_1)^2-m_{2}^{2}][(q+p_1+p_2)^2-m_{3}^{2}]
[(q+p_1+p_2+p_3)^2-m_{4}^{2}]\}^{-1}.
$$
In this appendix, we use the notations defined below for abbreviation:
\begin{eqnarray*}
B_0^{(1)},B_1^{(1)}&=&B_0,B_1\left[-k_1, m_{{\tilde{d}}^{I}_{i}}, m_{l^{J}}\right]\\
B_0^{(2)},B_1^{(2)}&=&B_0,B_1\left[-k_1, m_{{\tilde{l}}^{J}_{i}}, m_{d^{I}}\right]\\
B_0^{(3)},B_1^{(3)}&=&B_0,B_1\left[-k_2, m_{{\tilde{d}}^{I}_{i}}, m_{l^{J}}\right]\\
B_0^{(4)},B_1^{(4)}&=&B_0,B_1\left[-k_2, m_{{\tilde{l}}^{J}_{i}}, m_{d^{I}}\right]\\
B_0^{(5)},B_1^{(5)}&=&B_0,B_1\left[k_1-p_1, m_{{\tilde{d}}^{I}_{i}}, m_{l^{J}}\right]\\
B_0^{(6)},B_1^{(6)}&=&B_0,B_1\left[k_1-p_1, m_{{\tilde{l}}^{J}_{i}}, m_{d^{I}}\right]\\
B_0^{(7)}&=&B_0\left[p_1, m_{d^{I}}, m_{d^{I}}\right]\\
C_0^{(1)},C_{ij}^{(1)}&=&C_0,C_{ij}\left[-k_1, p_1, l^{J}, m_{{\tilde{d}}^{I}_{i}},
 m_{{\tilde{d}}^{I}_{i}}\right]\\
C_0^{(2)},C_{ij}^{(2)}&=&C_0,C_{ij}\left[-k_1, p_1, m_{{\tilde{l}}^{J}_{i}}, m_{d^{I}},
m_{d^{I}}\right]\\
C_0^{(3)},C_{ij}^{(3)}&=&C_0,C_{ij}\left[k_1, -p_1-p_2, m_{l^{J}}, m_{{\tilde{d}}^{I}_{i}},
 m_{{\tilde{d}}^{I}_{i}}\right]\\
C_0^{(4)},C_{ij}^{(4)}&=&C_0,C_{ij}\left[k_1, -p_1-p_2, m_{{\tilde{l}}^{J}_{i}}, m_{d^{I}},
m_{d^{I}}\right]\\
C_{ij}^{(5)}&=&C_0,C_{ij}\left[-p_2, k_1-p_1, m_{{\tilde{d}}^{I}_{i}},
m_{{\tilde{d}}^{I}_{i}}, m_{l^{J}}\right]\\
C_0^{(6)},C_{ij}^{(6)}&=&C_0,C_{ij}\left[-p_2, k_1-p_1, m_{d^{I}}, m_{d^{I}},
m_{{\tilde{l}}^{J}_{i}}\right]\\
C_0^{(7)},C_{ij}^{(7)}&=&C_0,C_{ij}\left[k_2, k_1, m_{{\tilde{d}}^{I}_{i}}, m_{l^{J}},
 m_{{\tilde{d}}^{I}_{i}}\right]\\
D_0^{(1)},D_{ij}^{(1)},D_{ijk}^{(1)}&=&D_0,D_{ij},D_{ijk}\left[k_1, -p_1, -p_2, m_{l^{J}},
m_{{\tilde{d}}^{I}_{i}}, m_{{\tilde{d}}^{I}_{i}}, m_{{\tilde{d}}^{I}_{i}}\right]\\
D_0^{(2)},D_{ij}^{(2)},D_{ijk}^{(2)}&=&D_0,D_{ij},D_{ijk}\left[k_1, -p_1, -p_2, m_{{\tilde{l}}^{J}_{i}},
 m_{d^{I}}, m_{d^{I}}, m_{d^{I}}\right]\\
F^{V}&=& -V_{{\tilde{d}}^{I}_{i} l^{J} c}^{L*} V_{{\tilde{d}}^{I}_{i} l^{J} t}^{L}\\
E^{V}&=& V_{d^{I} c \tilde{l}^{J}_{i}}^{R} V_{d^{I} t \tilde{l}^{J}_{i}}^{R*}\\
\Denoma &=&  \frac{1}{\hat{s}}\\
\Denomb =  \frac{1}{k_1^2-m_c^2}  && \Denomc =  \frac{1}{k_2^2-m_t^2}\\
\Denomd =  \frac{1}{\hat{t}-m_c^2}&& \Denome =  \frac{1}{\hat{t}-m_t^2}\\
\end{eqnarray*}
where the upper and lower indexes $I$, $J$ and $K$ appearing in
above variables denote the generation numbers $(I,J,K=1,2,3)$, and
lower indexes $i$ appearing in the supersymmetric quarks
$(\tilde{u}_i)$, $(\tilde{d}_i)$ and lepton $(\tilde{l}_i)$ can be
1 and 2.
\par
We use the denotation ${\cal T}$ in below to represent the
replacement of $(E^V \rightarrow F^V, m_{{\tilde{l}}^{J}_{i}}
\rightarrow m_{{\tilde{d}}^{I}_{i}}, m_{d^{I}} \rightarrow
m_{l^{J}}$) for the terms appearing before ${\cal T}$ in the same
level parentheses. We listed the expressions of $f_1$ to $f_{10}$
only and the others can obtained the transformation,
$f_{i+10}=-f_i(m_t \rightarrow -m_t),i=1\sim10$. The
factors $f_i$ we don't mention below, are zero.
\par
The form factors of the amplitude part from t-channel box
diagrams are written as
\begin{eqnarray*}
f_1^{b,\hat{t}} &=& \frac{i g_{s}^2}{8 \pi^2} \left\{ E^V \left[-
D_{313}^{(2)} m_c + ( - D_{311}^{(2)}  + D_{313}^{(2)}) m_t \right]+{\cal T} - E^V
D_{27}^{(2)} m_t \right\}\\
f_2^{b,\hat{t}} &=& \frac{i g_{s}^{2}}{32 \pi^2}  E^V \left[ (2
D_{27}^{(2)} + 6 D_{313}^{(2)} ) m_c+ (- D_{13}^{(2)}-D_{25}^{(2)}- D_{37}^{(2)}-
D_{23}^{(2)})m_c^3 \right.\\  &+& \left.(4 D_{27}^{(2)}+ 6 D_{311}^{(2)}- 6
D_{313}^{(2)}) m_t + (D_{23}^{(2)}-D_{25}^{(2)}- D_{35}^{(2)}+ D_{37}^{(2)}) m_c^2 m_t
\right.
\\ &+& \left.(-D_0^{(2)}- 2 D_{11}^{(2)}+D_{12}^{(2)}- D_{21}^{(2)}+ D_{24}^{(2)}
-D_{25}^{(2)}+D_{310}^{(2)}- D_{35}^{(2)} + D_{26}^{(2)}) m_c
m_t^2 \right.\\ &+&\left.
 (- D_{21}^{(2)}+ D_{24}^{(2)}+ D_{25}^{(2)}- D_{26}^{(2)}- D_{310}^{(2)}- D_{31}^{(2)}
 + D_{34}^{(2)}+D_{35}^{(2)} ) m_t^3 \right. \\ &+& \left.
 (D_{25}^{(2)}+ D_{37}^{(2)}- D_{26}^{(2)}- D_{39}^{(2)}) m_c \hat{s} +
 ( D_{35}^{(2)}-  D_{37}^{(2)} -D_{310}^{(2)} +D_{39}^{(2)}  ) m_t \hat{s}
 \right.\\&+& \left.
(- D_{12}^{(2)}- D_{24}^{(2)}-D_{310}^{(2)}+D_{37}^{(2)}+ D_{23}^{(2)}- D_{26}^{(2)} )m_c
\hat{t}+
 (D_0^{(2)} + D_{13}^{(2)})  m_c m_{d^{I}}^2 +
 (D_{11}^{(2)}- D_{13}^{(2)})  m_t m_{d^{I}}^2 \right. \\ &+& \left.
 (- D_{11}^{(2)}+ D_{13}^{(2)}- D_{21}^{(2)}- D_{23}^{(2)}- D_{24}^{(2)}
 + 2 D_{25}^{(2)}+ D_{26}^{(2)}+
  D_{310}^{(2)}-  D_{34}^{(2)}+ D_{35}^{(2)}- D_{37}^{(2)}) m_t \hat{t} \right]\\
f_3^{b,\hat{t}} &=& \frac{i g_{s}^2}{8 \pi^2} \left\{ E^V
\left[(D_{13}^{(2)} + 2 D_{25}^{(2)} +  D_{35}^{(2)} ) m_c +  (D_{11}^{(2)} \right.\right. \\
&-& \left.\left. D_{13}^{(2)}+ 2 D_{21}^{(2)}- 2 D_{25}^{(2)}+  D_{31}^{(2)}-
D_{35}^{(2)})
m_t\right]+{\cal T} \right\}\\
f_4^{b,\hat{t}} &=& \frac{i g_{s}^2}{16 \pi^2}  \left\{ E^V \left[
-2 D_{311}^{(2)}+ 6 D_{313}^{(2)}+(-D_{13}^{(2)}-D_{25}^{(2)}-D_{37}^{(2)}- D_{23}^{(2)})
m_c^2 \right.\right.\\
  &+& \left.\left. (D_0^{(2)}  + 2 D_{11}^{(2)}  +  D_{21}^{(2)} ) m_c m_t
  + (- D_{25}^{(2)}+D_{310}^{(2)}-D_{35}^{(2)} + D_{26}^{(2)}) m_t^2
  +( D_{25}^{(2)}+ D_{37}^{(2)} - D_{26}^{(2)} \right.\right.\\
  &-&\left.\left. D_{39}^{(2)}) \hat{s} +
 (-D_{310}^{(2)}+ D_{37}^{(2)} - D_{26}^{(2)}+ D_{23}^{(2)}) \hat{t}
 + (D_0^{(2)}+  D_{13}^{(2)}) m_{d^{I}}^2\right] +2 F^V (-D_{27}^{(1)} -  D_{311}^{(1)})
     \right\}\\
f_5^{b,\hat{t}} &=& \frac{i g_{s}^2}{16 \pi^2} \left\{
 2 E^V ( D_{27}^{(2)} + D_{311}^{(2)})-{\cal T} + E^V \left[2 D_{311}^{(2)}
  - 6 D_{313}^{(2)} \right. \right.\\
 &+& \left.\left. (D_{23}^{(2)}  - D_{25}^{(2)}  -  D_{35}^{(2)}  +  D_{37}^{(2)}  )  m_c^2+
  ( - D_{11}^{(2)}  + D_{13}^{(2)}  -  2 D_{21}^{(2)}  + D_{24}^{(2)} + 2 D_{25}^{(2)}  -
 D_{26}^{(2)} \right.\right.\\ &-& \left.\left. D_{310}^{(2)}  -  D_{31}^{(2)}
 +D_{34}^{(2)}+ D_{35}^{(2)})m_t^{(2)}
 +( - D_{310}^{(2)}   + D_{39}^{(2)}+  D_{35}^{(2)} -  D_{37}^{(2)} ) \hat{s} \right. \right.\\
 &+& \left.\left. (D_{35}^{(2)} -  D_{37}^{(2)}- D_{23}^{(2)}  - D_{24}^{(2)}  + D_{25}^{(2)}
 + D_{26}^{(2)}+D_{310}^{(2)}
  - D_{34}^{(2)}   )  \hat{t} + (D_{11}^{(2)}  - D_{13}^{(2)} ) m_{d^{I}}^2\right]\right\}\\
f_6^{b,\hat{t}} &=& \frac{i g_{s}^2}{8 \pi^2} \left[ E^V(
D_{312}^{(2)}-  D_{313}^{(2)}) +{\cal T} + E^V D_{27}^{(2)} \right]\\
f_7^{b,\hat{t}} &=& \frac{i g_{s}^2}{32 \pi^2}
 E^V \left[ -2 D_{27}^{(2)}- 6 D_{312}^{(2)} + 6 D_{313}^{(2)}+
 (D_0^{(2)}+D_{11}^{(2)} )m_c m_t\right. \\ &+&\left.
 (- D_{13}^{(2)}-D_{25}^{(2)}+D_{310}^{(2)}-D_{37}^{(2)}-D_{23}^{(2)}+D_{26}^{(2)}) m_c^2
 \right.\\&+&\left.
 (-D_{11}^{(2)}+D_{12}^{(2)}-D_{21}^{(2)}-D_{22}^{(2)}+ 2 D_{24}^{(2)}-D_{25}^{(2)}
 +D_{310}^{(2)}+D_{34}^{(2)}-D_{35}^{(2)}-D_{36}^{(2)} +D_{26}^{(2)}) m_t^2 \right.\\&+&\left.
 (+D_{25}^{(2)}-D_{310}^{(2)}+D_{37}^{(2)}-D_{26}^{(2)}+D_{38}^{(2)}-D_{39}^{(2)} )\hat{s} +
 (D_{22}^{(2)}-2 D_{26}^{(2)}- 2 D_{310}^{(2)}\right. \\&+&\left.
D_{36}^{(2)}+D_{37}^{(2)}+D_{23}^{(2)} )\hat{t}+ (D_0^{(2)}-D_{12}^{(2)}+D_{13}^{(2)})
m_{d^{I}}^2 \right]\\
f_8^{b,\hat{t}} &=& \frac{i g_{s}^2}{8 \pi^2}\left[
 E^V (-D_{24}^{(2)}+ D_{25}^{(2)}- D_{34}^{(2)}+  D_{35}^{(2)})+{\cal T}
 +F^V (-D_{12}^{(1)}+ D_{13}^{(1)}-  D_{24}^{(1)}+   D_{25}^{(1)}) \right]\\
f_9^{b,\hat{t}} &=& \frac{i g_{s}^2}{16 \pi^2}  E^V \left[
D_{12}^{(2)}+ D_{24}^{(2)} ) m_c + (D_{11}^{(2)}- D_{12}^{(2)}+
D_{21}^{(2)}- D_{24}^{(2)} ) m_t \right] \\
f_{10}^{b,\hat{t}} &=& \frac{i g_{s}^{2}}{16 \pi^2}  E^V \left[
(-D_{13}^{(2)}-D_{25}^{(2)})  m_c +
(-D_{11}^{(2)}+D_{13}^{(2)}-D_{21}^{(2)}+D_{25}^{(2)})  m_t\right] \\
\end{eqnarray*}

\par
The form factors of the amplitude part from t-channel vertex
diagrams are written as
\begin{eqnarray*}
f_2^{v,\hat{t}} &=& \frac{i g_{s}^2}{64 \pi^2} \left\{ 2 \Denome
  (\hat{t}-m_t^2) C_{12}^{(6)} E^V m_c  + \Denomd
 \left[E^V( (m_c - m_t)( 1 - 4 C_{24}^{(2)} - 2 C_0^{(2)}  m_{d^{I}}^{(2)})\right.\right.\\
  &+& \left.\left.2( C_{11}^{(2)} +C_{21}^{(2)}) m_t^2+ 2(C_{12}^{(2)}
  +C_{23}^{(2)})(\hat{t} -m_t^2) )
 - 2 (C_{11}^{(2)}+  C_0^{(2)}) m_t (\hat{t}-m_t m_c) )\right.\right.\\
 &+&\left.\left.4 C_{24}^{(1)} F^V (m_c- m_t) \right]\right\} \\
f_3^{v,\hat{t}} &=& \frac{i g_{s}^2}{8 \pi^2}  \Denome \left\{
E^V\left[ (- C_{12}^{(6)} - C_{23}^{(6)}) m_c + (C_{23}^{(6)} -
C_{22}^{(6)}) m_t
\right]+{\cal T} \right]\\
f_4^{v,\hat{t}} &=& \frac{i g_{s}^2}{32 \pi^2} \left\{ \Denome
\left[E^V ( 1 - 4 C_{24}^{(6)} + 2(  C_{12}^{(6)}  +  C_{23}^{(6)}) m_c^2 + 2(
C_{22}^{(6)}- C_{23}^{(6)}) \hat{t}
 - 2 C_0^{(6)} m_{d^{I}}^2 + 2 C_{12}^{(6)} m_c m_t )\right.\right.\\
&+& \left.\left.4 C_{24}^{(5)} F^V\right]+2 \Denomd \left[ E^V (2
C_{24}^{(2)} +(C_{23}^{(2)}- C_{21}^{(2)}) m_t^2- C_{23}^{(2)} \hat{t}- (C_{11}^{(2)}
+ C_{21}^{(2)})m_t m_c )+{\cal T} \right.\right.\\&+&  \left. \left.E^V
\left[-B_0^{(7)} - C_0^{(2)} m_{d^{I}}^2 + C_0^{(2)}
m_{{\tilde{l}}^{J}_{i}}^2 - (C_0^{(2)} + C_{11}^{(2)})m_t m_c
\right]+F^V \left[(- C_{11}^{(1)} + C_{12}^{(1)})m_t^2 - C_{12}^{(1)}
\hat{t} \right]\right]\right\} \\
f_5^{v,\hat{t}} &=& \frac{i g_{s}^2}{16 \pi^2}  \Denome
\left[(C_{23}^{(6)}- C_{22}^{(6)}) E^V (\hat{t}- m_t^2) +{\cal T}\right]\\
f_7^{v,\hat{t}} &=& \frac{i g_{s}^2}{64 \pi^2} \left\{ \Denome
\left[ E^V ( 1 - 4 C_{24}^{(6)} + 2(  C_{12}^{(6)}  +
C_{23}^{(6)}) m_c^2 + 2( C_{22}^{(6)}- C_{23}^{(6)}) \hat{t}
 - 2 C_0^{(6)} m_{d^{I}}^2 + 2 C_{12}^{(6)} m_c m_t )\right.\right.\\
&+&4 C_{24}^{(5)} F^V \left]+\Denomd \left[ E^V (1 - 4 C_{24}^{(2)} + 2(
C_{11}^{(2)} - C_{12}^{(2)} +  C_{21}^{(2)}  - C_{23}^{(2)}) m_t^2 \right.\right.\\
&+& \left.\left. 2 (C_{12}^{(2)}  +  C_{23}^{(2)}) \hat{t} - 2 C_0^{(2)}
m_{d^{I}}^2 - 2(C_0^{(2)}+ C_{11}^{(2)} )m_t m_c ) +4 C_{24}^{(1)} F^V\right] \right\} \\
f_9^{v,\hat{t}} &=& \frac{i g_{s}^2}{16 \pi^2} \Denomd \left\{
E^V\left[(-C_{11}^{(2)}  + C_{12}^{(2)}  - C_{21}^{(2)}  + C_{23}^{(2)} )
m_t  +(- C_{12}^{(2)}  - C_{23}^{(2)})m_c  \right] +{\cal T}\right\}\\
f_{10}^{v,\hat{t}} &=& \frac{i g_{s}^2}{16 \pi^2}  \Denome
\left\{E^V\left[ ( C_{12}^{(6)} + C_{23}^{(6)}) m_c + (-C_{23}^{(6)} +
C_{22}^{(6)}) m_t\right]+{\cal T}\right\}\\
\end{eqnarray*}
\par
The form factors of the amplitude part from t-channel self-energy
diagrams are written as
\begin{eqnarray*}
f_2^{s,\hat{t}} &=& \frac{i g_{s}^2}{32 \pi^2} E^V
   \left[- \Denomb \Denomd (B_0^{(2)} + B_1^{(2)}) (m_t^2 - m_c^2) m_t+
      \Denomd \Denome (B_0^{(6)} + B_1^{(6)}) (\hat{t} - m_t^2) m_c\right]-{\cal T}\\
f_4^{s,\hat{t}} &=& \frac{i g_{s}^2}{16 \pi^2} E^V
   \left[\Denomb \Denomd (B_0^{(2)} + B_1^{(2)}) (m_t + m_c) m_t+
    \Denomc \Denome (B_0^{(4)} + B_1^{(4)}) (m_t + m_c) m_c\right.\\
   &+&\left.\Denomd \Denome (B_0^{(6)} + B_1^{(6)}) (\hat{t} + m_t m_c)\right]-{\cal T}\\
f_7^{s,\hat{t}} &=& \frac{i g_{s}^2}{32 \pi^2} E^V
   \left[\Denomb \Denomd (B_0^{(2)} + B_1^{(2)}) (m_t + m_c) m_t+
    \Denomc \Denome (B_0^{(4)} + B_1^{(4)}) (m_t + m_c) m_c\right.\\
    &+&\left. \Denomd \Denome (B_0^{(6)} + B_1^{(6)}) (\hat{t} + m_t m_c)\right]-{\cal T}\\
\end{eqnarray*}
The form factors of the amplitude part from s-channel
diagrams are written as
\begin{eqnarray*}
f_1^{\hat{s}} &=& \frac{i g_{s}^2}{64 \pi^2} \Denoma \left\{
     E^V\left[ ( -2 \Denomb (B_0^{(2)} + B_1^{(2)}) (m_t^2-m_c^2) m_t
     -2 \Denomc (B_0^{(4)} + B_1^{(4)}) (m_t^2-m_c^2) m_c -{\cal T}) \right.\right.\\
    &+&\left.\left.(m_c -  m_t)(1- 4 C_{24}^{(4)}  - 2 C_0^{(4)}   m_{d^{I}}^2)
 + 2( C_0^{(4)}   + 2 C_{11}^{(4)}   -  C_{12}^{(4)}+ C_{21}^{(4)} - 2 C_{23}^{(4)}
 +  C_{22}^{(4)} ) m_c m_t^2\right.\right. \\
 &+& \left.\left.2( C_{12}^{(4)}  +  C_{22}^{(4)} ) m_c^3
 + 2( C_{12}^{(4)}  +  2 C_{23}^{(4)}  -  C_{22}^{(4)} ) m_c \hat{t}
 + 2(- C_{12}^{(4)}    - _{22}^{(4)})  m_c \hat{u}\right.\right.\\
 &+& \left.\left.2( -  C_0^{(4)}  - C_{11}^{(4)} - C_{12}^{(4)} -  C_{22}^{(4)} ) m_c^2 m_t
 + 2(-  C_{11}^{(4)}  +  C_{12}^{(4)}  -  C_{21}^{(4)}   +  2 C_{23}^{(4)}  - C_{22}^{(4)} ) m_t^3\right.\right.\\
 &+& \left.\left.2( C_{11}^{(4)}  -  C_{12}^{(4)}  + C_{21}^{(4)}  -  2 C_{23}^{(4)}   +  C_{22}^{(4)})  m_t \hat{t}
 + 2(- C_{11}^{(4)} +   C_{12}^{(4)}  - C_{21}^{(4)}   +  C_{22}^{(4)} ) m_t \hat{u}\right]\right.\\ &+&\left.
   2 F^V \left[ 2 C_{24}^{(3)} (m_c -m_t)
+  (C_{12}^{(3)}+C_{23}^{(3)}) (m_c-m_t) (\hat{t} - \hat{u})
 +  (C_{11}^{(3)}+C_{21}^{(3)}) m_t (\hat{t} - \hat{u} )
  \right] \right\}  \\
f_6^{\hat{s}} &=& \frac{i g_{s}^2}{32 \pi^2} \Denoma  \left\{
             E^V\left[( 2 \Denomb (B_0^{(2)} + B_1^{(2)}) (m_t + m_c) m_t+
  2 \Denomc (m_t + m_c) (B_0^{(4)} + B_1^{(4)}) m_c-{\cal T})\right.\right.\\ &+&\left.\left.
  1  - 4 C_{24}^{(4)} + 2 (C_{12}^{(4)} + C_{23}^{(4)}) m_c^2 - 2 (C_0^{(4)}+C_{11}^{(4)}) m_c m_t
+ 2 (C_{11}^{(4)}+C_{21}^{(4)}-C_{12}^{(4)}- C_{23}^{(4)})m_t^2\right.\right.  \\
&+& \left.\left.2 (C_{22}^{(4)} - C_{23}^{(4)}) \hat{s} - 2 C_0^{(4)}
m_{d^{I}}^2\right]+ 4 C_{24}^{(3)} F^V  \right\}\\
\end{eqnarray*}
\par
The form factors of the amplitude part from quartic
diagram are written as
\begin{eqnarray*}
f_1^{q} &=& \frac{i g_{s}^2}{32 \pi^2}
F^V \left[ (-C_0^{(7)} - C_{11}^{(7)} ) m_c + C_{12}^{(7)} m_t \right]\\
\end{eqnarray*}
\end{large}

\vskip 10mm
\begin{flushleft} {\bf Figure Captions} \end{flushleft}

{\bf Fig.1} The Feynman diagrams of the subproecesses
$d\bar{d} \rightarrow t\bar{c}+\bar{t}c$
and $gg \rightarrow t\bar{c}+\bar{t}c$.

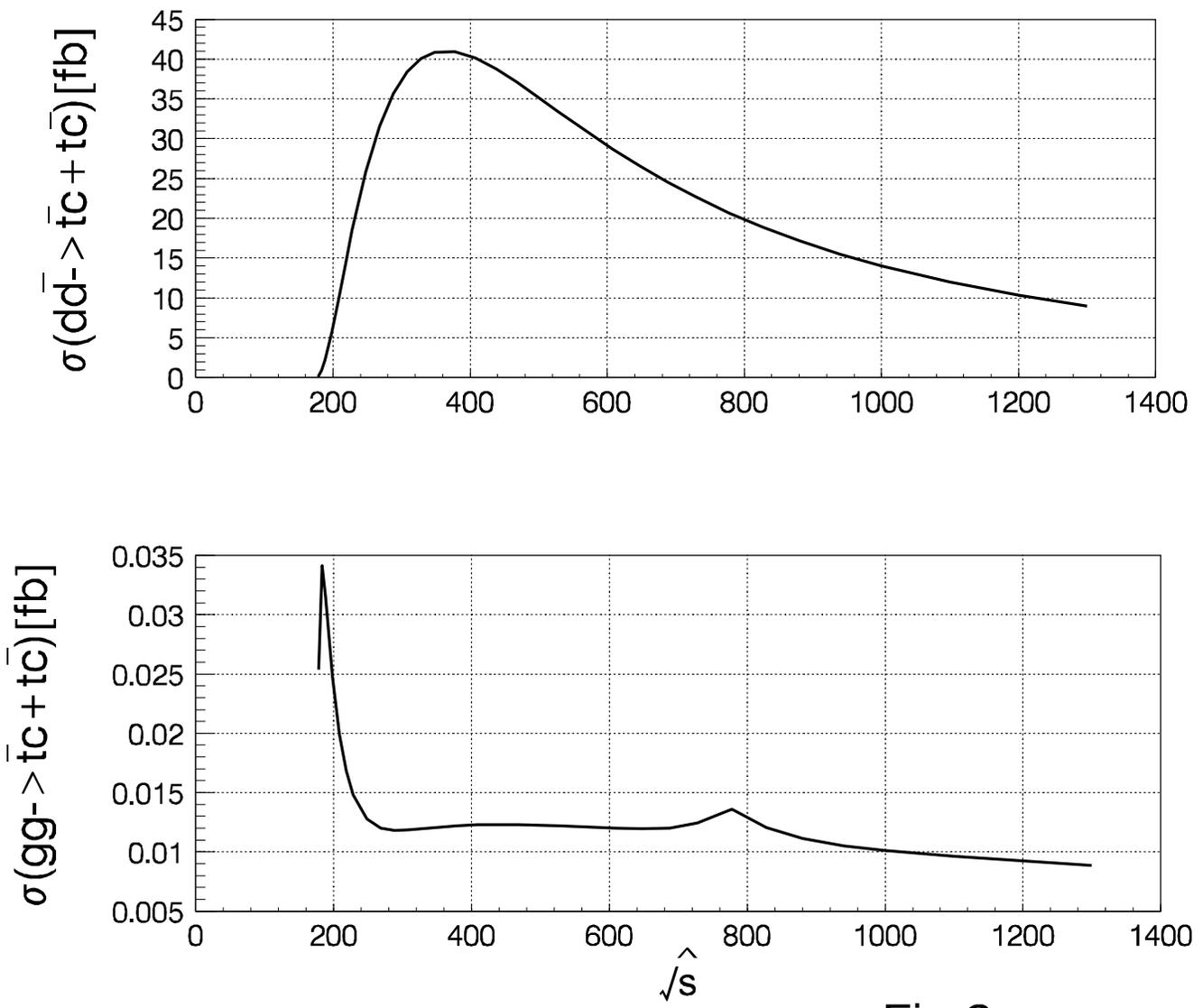
{\bf Fig.2} The subprocess cross sections as a function of
$\sqrt{\hat{s}}$. The upper is of $d\bar{d} \rightarrow
t\bar{c}+\bar{t}c$ and the lower is of $gg \rightarrow
t\bar{c}+\bar{t}c$.

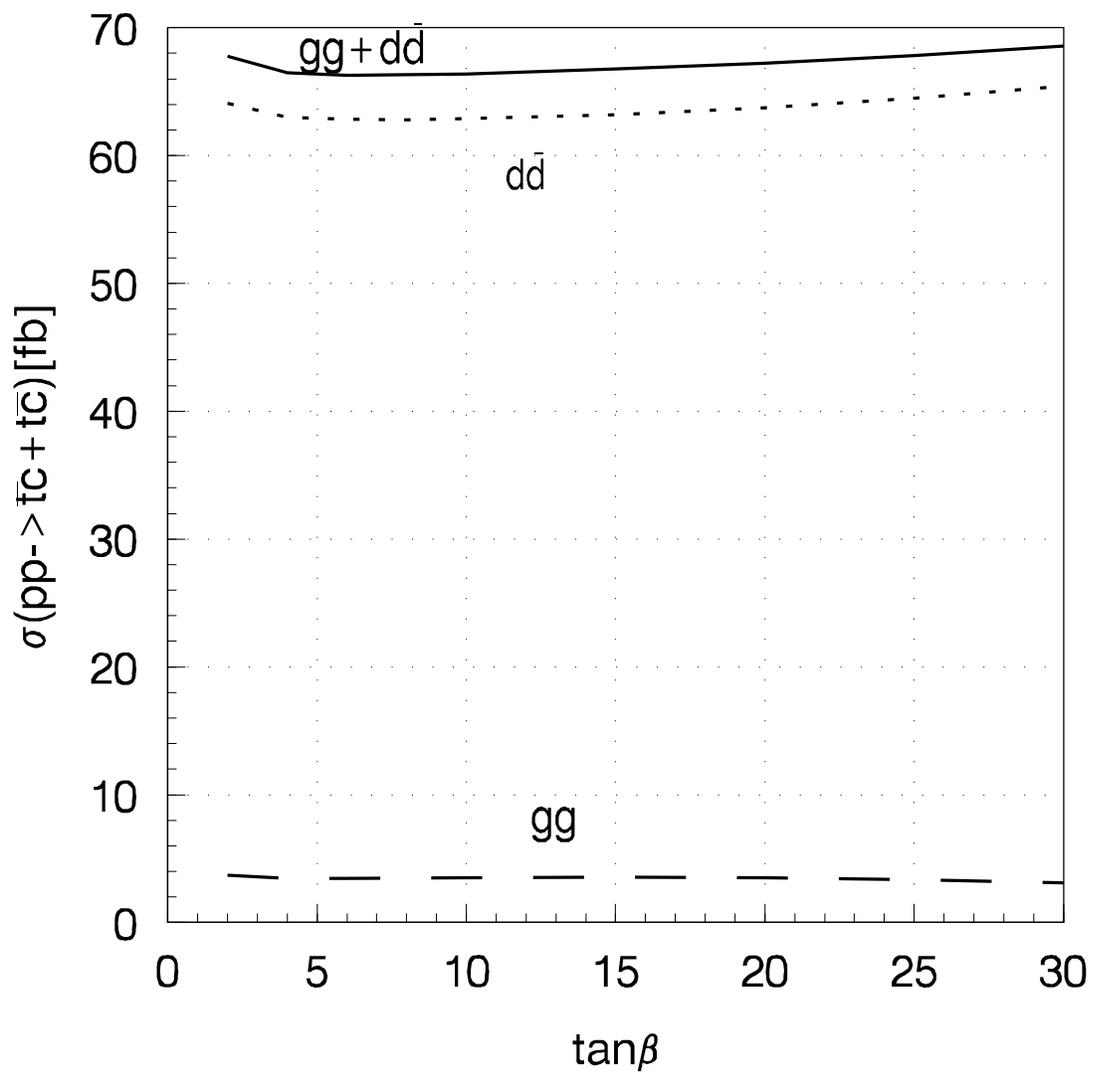
{\bf Fig.3} The folded cross sections as a function of $\tan\beta$
at LHC in the mSUGRA scenario.

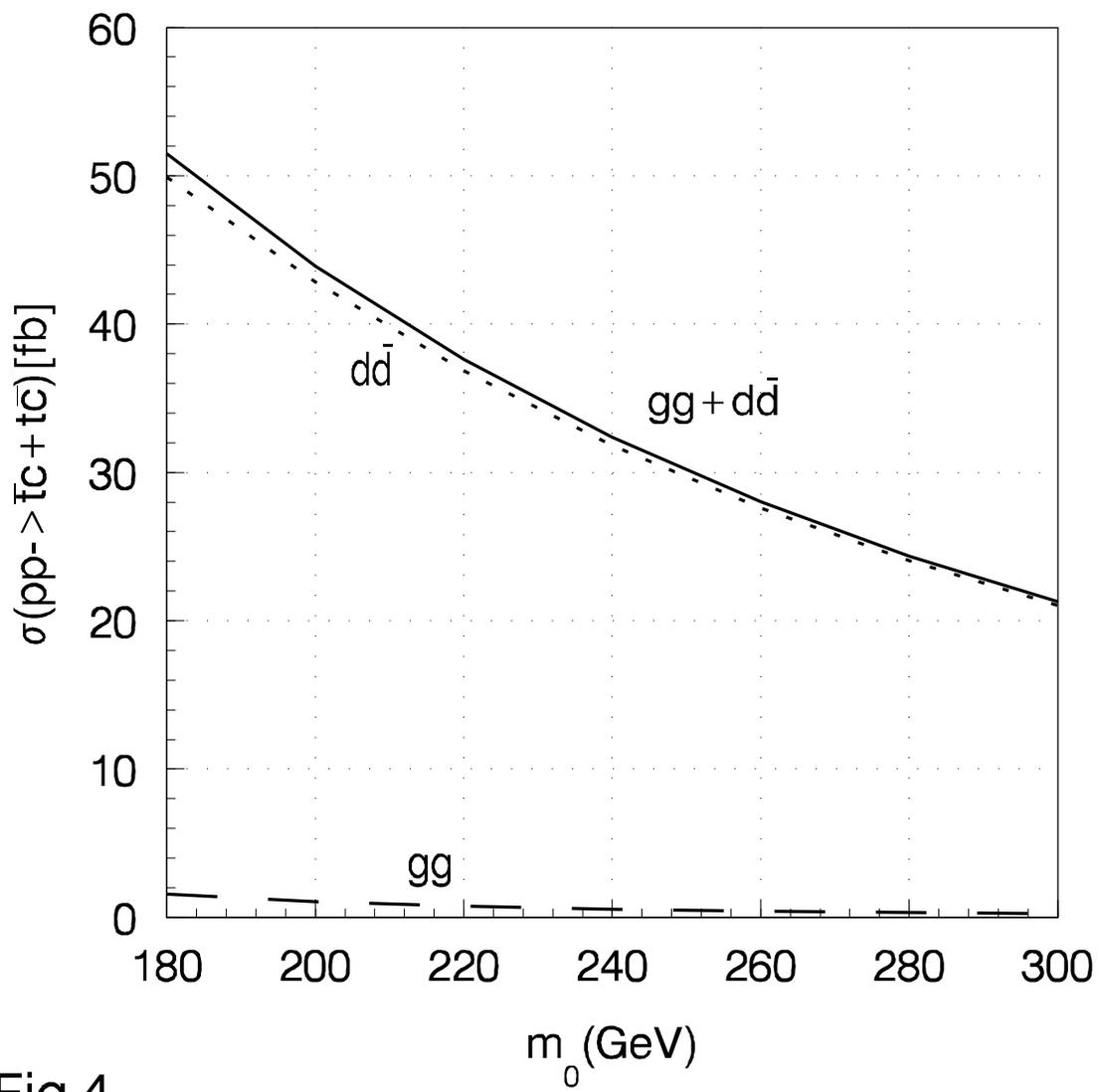
{\bf Fig.4} The folded cross sections as a function of $m_0$ at
LHC in the mSUGRA scenario.

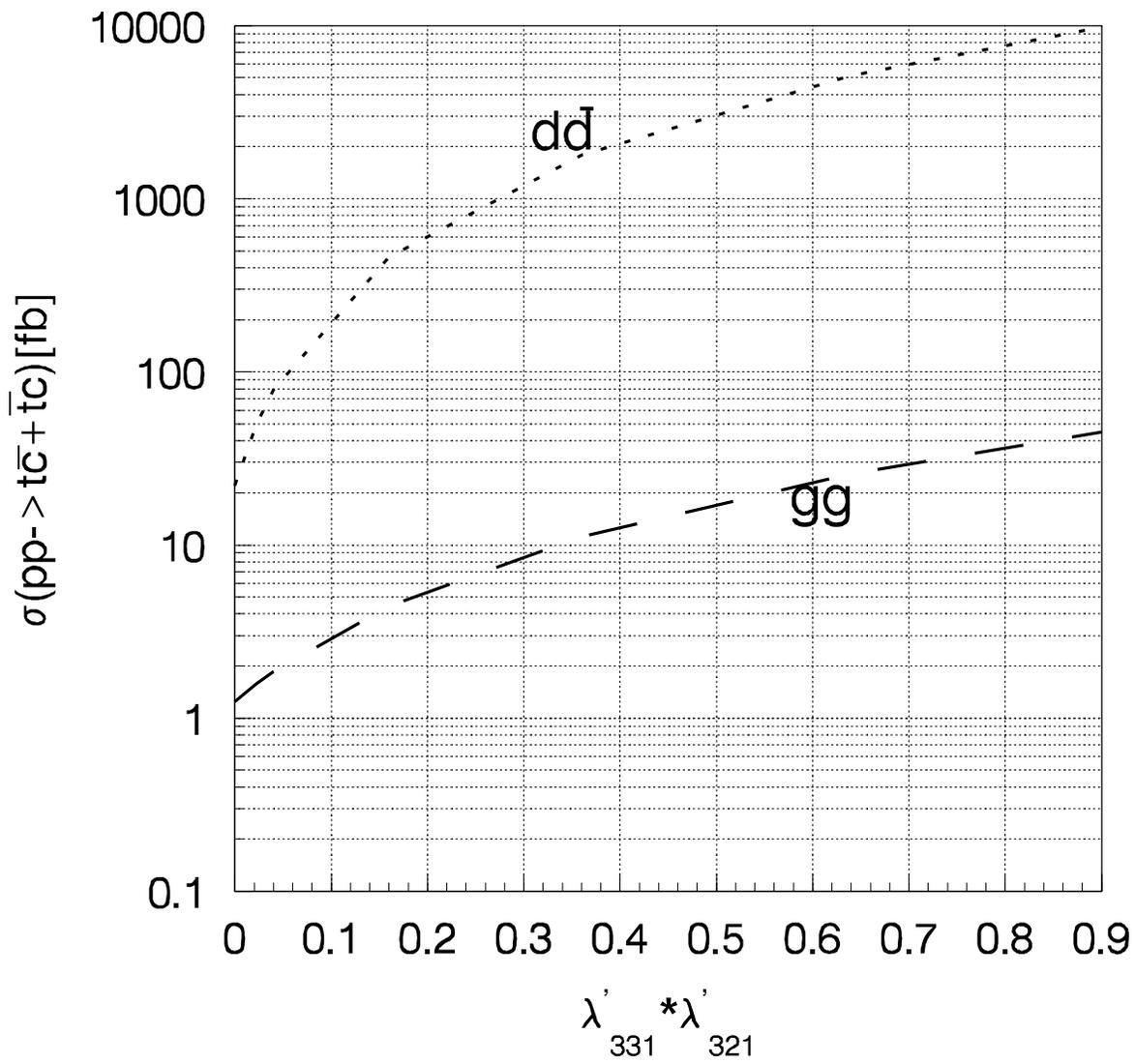
{\bf Fig.5} The folded cross sections as a function of $\lambda^{'}_{331}* \lambda^{'}_{321}$
at LHC.
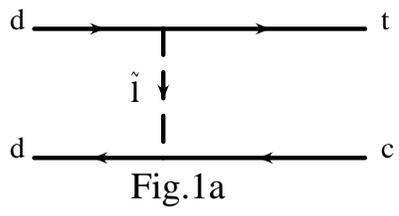
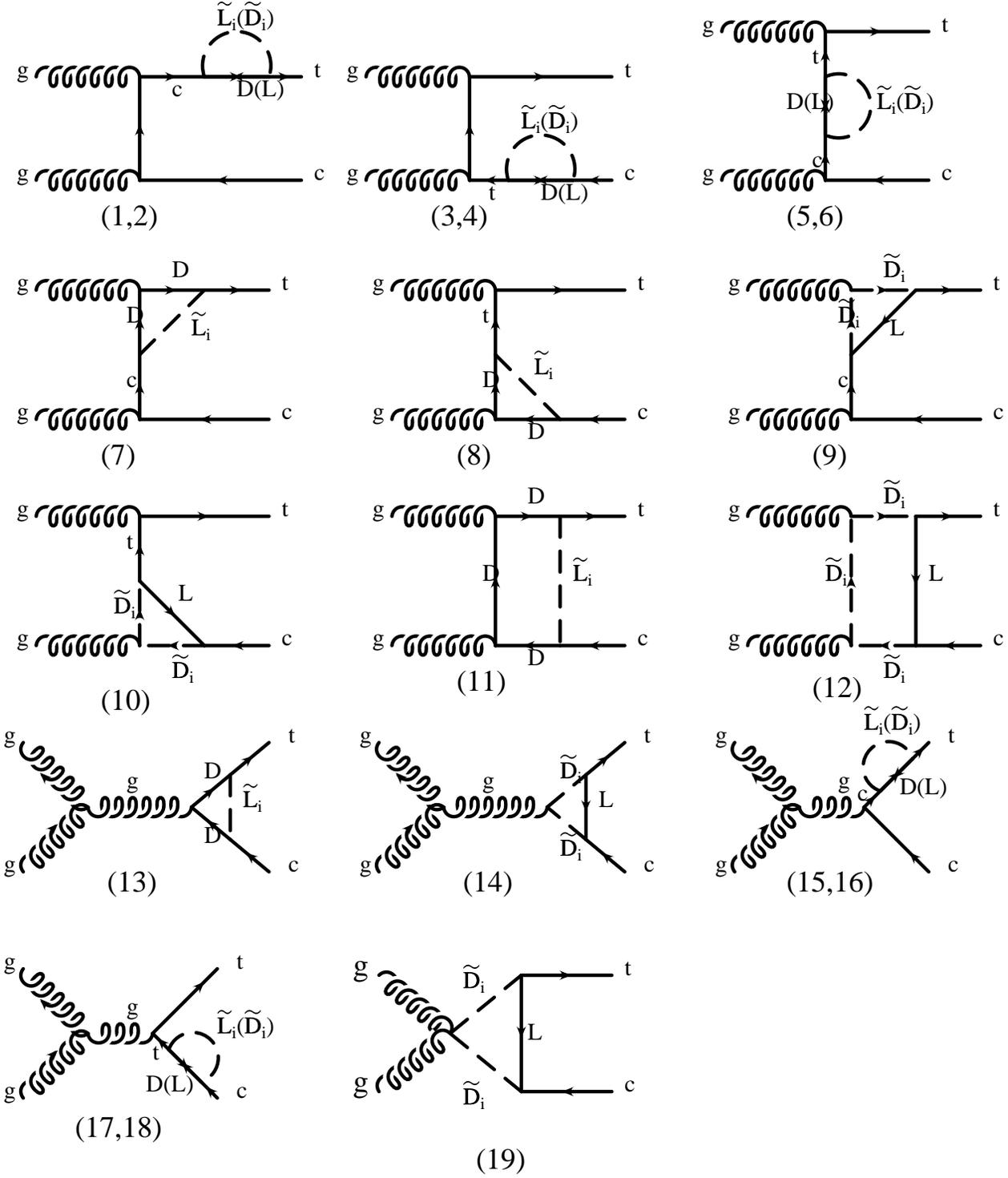
\end{document}